\documentclass[12pt]{article}

\advance \textheight by \topskip
\makeatletter
\@addtoreset{equation}{section}
 
\makeatother

\def\R{\mathbb R}

\def\X{\mathfrak X}
\def\P{\mathfrak P}
\def\K{\mathfrak K}
\def\J{\mathfrak J}
\def\D{\mathfrak D}
\def\mH{\mathfrak H}

\usepackage{amsmath,amssymb}

\begin{document}
\title{
\begin{flushright}
{\small USACH-FM-03/02}\\[1.0cm]
\end{flushright}
{\bf Conformal Symmetry of
Relativistic and Nonrelativistic
Systems and AdS/CFT Correspondence}}

\author{{\sf Carlos Leiva${}^{a}$}\thanks{
E-mail: caleiva@lauca.usach.cl}
{\sf\ and Mikhail S. Plyushchay${}^{a,b}$}\thanks{
E-mail: mplyushc@lauca.usach.cl}
\\
{\small {\it ${}^a$Departamento de F\'{\i}sica,
Universidad de Santiago de Chile,
Casilla 307, Santiago 2, Chile}}\\
{\small {\it ${}^b$Institute for High Energy Physics,
Protvino, Russia}}}
\date{}






\maketitle


\begin{abstract}
The nonlinear realization
of conformal $so(2,d)$ symmetry for relativistic
systems  and  the dynamical  conformal $so(2,1)$ symmetry
of nonrelativistic  systems
are investigated  in the context of AdS/CFT correspondence.
We show that the massless particle
in $d$-dimensional Minkowski space can be treated
as the system confined to the border
of the AdS${}_{d+1}$ of infinite radius,
while various nonrelativistic systems
may be canonically related
to a relativistic
(massless, massive, or tachyon) particle
on the AdS${}_{2}\times$S${}^{d-1}$.
The list of nonrelativistic systems ``unified"
by such a correspondence comprises
the conformal mechanics model,
the planar charge-vortex and
$3$-dimensional
charge-monopole systems,
the particle in a planar
gravitational field of a point massive source,
and the conformal model associated with the charged
particle propagating near the horizon of the
extreme Reissner-Nordstr\"om
black hole.
\end{abstract}

\bigskip
\section{\protect\bigskip Introduction}


Being a nontrivial
generalization of
the Poincar\'e symmetry, nowadays conformal symmetry
is of ever-increasing interest.
In the simplest form it appears as a
rigid $so(2,d)$
symmetry  of massless particles
and fields in $d$-dimensional Minkowski space-time.
This symmetry is also enjoyed by many
nonrelativistic quantum mechanical
\cite{AFF}-\cite{dgh}
and field theories \cite{Hag}-\cite{OR}
in the form of the unexpected $so(2,1)$ symmetry
sometimes called dynamical or hidden
(Schr\"odinger ) symmetry.
The recent revival of interest in conformal mechanics
is in the context of AdS/CFT correspondence
\cite{Mald,Gubs,WitE} (for review see \cite{Ahar})
and black hole physics \cite{kallosh}-\cite{Bell}.

It is long known that the
conformal $so(2,1)$ symmetry of nonrelativistic
quantum mechanical systems
may be understood  as a part
of the usual conformal $so(2,d)$ symmetry
which survives a tricky nonrelativistic
contraction procedure
applied to a corresponding relativistic
system \cite{BaRa}
(see also \cite{dgh,hh} for the Kaluza-Klein type approach).
On the other hand,
the AdS/CFT correspondence, or the ``holographic principle"
relates certain theories
on $d+1$ dimensional AdS with conformal field theories in
Minkowski space-time of one dimension less.
Therefore, the natural question arises as to whether the
dynamical conformal symmetry of nonrelativistic systems
can be somehow related to the conformal symmetry
of certain relativistic systems in the
context of AdS/CFT correspondence.
The purpose of the present work is to response
this question as well as to trace out in detail
how the nonlinear realization of the conformal symmetry
emerges in the same
context of AdS/CFT duality.

More specifically,
first, we shall investigate how
in the simplest case of the massless scalar particle
its conformal symmetry
in $d$-dimensional Minkowski space
originates from the Lorentz symmetry
of the corresponding ambient Minkowski space
having two more dimensions.
This will be done with the help of
the $d+2$-dimensional model
possessing the local (gauge) $so(2,1)$ conformal symmetry.
Then we shall reinterpret the model
as a massless particle
confined to the border of AdS${}_{d+1}$
of infinite radius.
Secondly, we shall demonstrate how a free nonrelativistic
massive particle in one dimension may be canonically related
to the massless particle on AdS${}_{2}$
of finite radius.
Generalizing this observation,
we shall construct
the Lagrangian for the
relativistic particle on the
AdS${}_{2}\times$S${}^{d-1}$,
which is related to the known
nonrelativistic particle systems
enjoying the dynamical $so(2,1)$ symmetry.
The corresponding list
will include the $d$-dimensional conformal mechanics model
\cite{AFF}, the planar charge-vortex and $3$-dimensional
charge-monopole systems \cite{Jackm,Jackv},
the nonrelativistic particle in a planar
gravitational field \cite{Des},
and the
``new" conformal
mechanics model corresponding to the charged
particle propagating near the horizon of the
extreme Reissner-Nordstr\"om
black hole \cite{kallosh}.

The paper is organized as follows.
In Section 2 we review the conformal symmetry
of the massless particle in $d$-dimensional Minkowski
space-time, while Section 3 is devoted
to the discussion of the dynamical
conformal $so(2,1)$
symmetry of various mechanical
systems.  In Section 4 we show how
the nonlinearly realized conformal symmetry of the massless
particle in $d$ dimensions can be understood
in the context of AdS/CFT correspondence.
In Section 5 the dynamical
conformal symmetry of non-relativistic
mechanical systems is analyzed from the perspective
of AdS/CFT duality.
The obtained results are summarized in Section 6.

\section{\protect\bigskip Conformal Symmetry of Massless
Particle}

The infinitesimal transformations
\begin{equation}
\delta x^\mu = \omega^\mu{}_\nu x^\nu  + \alpha^\mu + \beta
x^\mu +
2(x\gamma)x^\mu - x^2\gamma^\mu
\label{inf1}
\end{equation}
generate the conformal symmetry,
$ds^2\rightarrow ds'{}^2=e^{2\sigma}ds^2$,
on the
$d$-dimensional Minkowski space $\R^{1,d-1}$
with metric
$$
ds^2=dx^\mu dx^\nu\eta_{\mu\nu}=-dx_0^2+\sum_{i=1}^{d-1}dx^{
2}_i.
$$
Here
the parameters
$\omega^\mu{}_\nu$,
$\alpha^\mu$,
$\beta$
and $\gamma^\mu$ correspond
to the Lorentz rotations,
space-time translations,
scale (dilatation) and special conformal transformations.
Due to a nonlinear (quadratic) in $x^\mu$
nature of the two  last terms in (\ref{inf1}),
the finite version of the special
conformal transformations,
\begin{equation}
x'{}^\mu=\frac{x^\mu-\gamma^\mu x^2}{1
-2\gamma x + \gamma^2 x^2},
\label{confin}
\end{equation}
is not defined globally\footnote{
Transformation (\ref{confin}) is singular at
$x^\mu=\gamma^\mu/\gamma^2$ when
$\gamma^2\neq 0$,
and at $x\gamma=1/2$
for $\gamma^2=0$.},
and to be well defined requires a
compactification of $\R^{1,d-1}$
by including the points at infinity
(for the discussion of the aspects of globality etc.,
see \cite{HE,WitE,Ahar}).

On the classical phase space with
canonical Poisson bracket relations
$\{x_\mu,$ $p_\nu\}=\eta_{\mu\nu}$,
$\{x_\mu,x_\nu\}=\{p_\mu,p_\nu\}=0$,
the transformations
(\ref{inf1}) are generated by
\begin{equation}
M_{\mu \nu } = x_{\mu }p_{\nu }-x_{\nu }p_{ \mu }, \qquad
P_{\mu } = p_{\mu },\qquad
D = x^{\mu }p_{\mu }, \qquad
K_{\mu } = 2x_{\mu }(xp)-x^{2}p_{\mu }.
\label{genconf}
\end{equation}
The generators (\ref{genconf})
form the conformal algebra
\begin{eqnarray}
&\{M_{\mu\nu},M_{\sigma\lambda}\}=
\eta_{\mu\sigma}M_{\nu\lambda}-
\eta_{\nu\sigma}M_{\mu\lambda}+
\eta_{\mu\lambda}M_{\sigma\nu}-
\eta_{\nu\lambda}M_{\sigma\mu},&\notag\\
&\{M_{\mu \nu },P_{\lambda }\} =
\eta_{\mu \lambda }P_{ \nu }-
\eta_{\nu\lambda }P_{\mu },\qquad
\{M_{\mu \nu },K_{\lambda }\} =
\eta_{\mu \lambda }K_{ \nu }-
\eta_{\nu\lambda }K_{\mu },&\notag\\
&\{D, P_{\mu }\} =P_{\mu },\qquad
\{D, K_{\mu }\} =-K_{\mu },&\notag\\
&\{ K_{\mu },P_{\nu }\} =
2(\eta_{\mu \nu }D+M_{\mu \nu}),&
\notag \\
&\{D, M_{\mu\nu}\} =
\{P_\mu,P_\nu\}=
\{ K_{\mu },K_{\nu }\} =0.&
\label{confalg}
\end{eqnarray}
The algebra (\ref{confalg})
is isomorfic to the algebra $so(2,d)$, and
by defining
\begin{equation}
J_{\mu \nu }=M_{\mu \nu },\quad
 J_{\mu d}=\frac{1}{2}(P_\mu + K_\mu),\quad
 J_{\mu (d+1)}=\frac{1}{2}(P_\mu-K_\mu),
 \quad
 J_{d(d+1)}=D,
 \label{ident}
\end{equation}
can be put in the standard form
\begin{equation}
\{J_{AB},J_{LN}\}=
\eta_{AL}J_{BN}-
\eta_{BL}J_{AN}+
\eta_{AN}J_{LB}-
\eta_{BN}J_{LA}
\label{so24}
\end{equation}
with $A,B=0,1,\ldots, d,d+1$, and
\begin{equation}
\eta_{AB}=diag (-1,+1,\ldots,+1,-1).
\label{etaab}
\end{equation}

The phase space constraint
$
\varphi_m\equiv p^2+m^2=0
$
describing the
free relativistic particle of mass $m$
in $\R^{1,d-1}$ is invariant under the Poincar\'e
transformations, $\{M_{\mu\nu},\varphi_m\}=
\{P_\mu,\varphi_m\}=0$.
Unlike the $P_\mu$ and $M_{\mu\nu}$,
the generators
of the scale and special conformal transformations
commute weakly with $\varphi_m$ only in the
massless case $m=0$:
$\{D,\varphi_0\}=2\varphi_0=0$,
$\{K_\mu,\varphi_0\}=4x_\mu\varphi_0=0$.
This means that $D$ and $K_\mu$ are the integrals of
motion of the free scalar massless particle
whose Lagrangian is
\begin{equation}
L=\frac{\dot{x}^2}{2e},
\label{m0}
\end{equation}
where $\dot x_\mu=\frac{d}{d\tau}x_\mu$,
and $e=e(\tau)$ is a Lagrange multiplier.
Lagrangian (\ref{m0}) is
invariant under the Poincar\'e transformations
given by Eq. (\ref{inf1}) with
$\beta=\gamma_\mu=0$ and supplied with the
relations $\delta_\omega e=\delta_\alpha e=0$.
The scale and special conformal symmetry
transformations act nontrivially
on the both $x_\mu$ and $e$.
The finite dilatations are
\begin{equation}
x'^\mu=\exp\beta \cdot x^\mu,
\qquad
e'=\exp 2\beta \cdot e,
\label{findil}
\end{equation}
and the
special conformal
transformations are given by
Eq. (\ref{confin})
and
\begin{equation}
e'=\frac{e}{(1-2x\gamma+\gamma^2 x^2)^2}.
\label{efin}
\end{equation}
Lagrangian (\ref{m0})
is also invariant under the inversion
transformation\footnote{Transformation (\ref{-1})
is well defined on the compactified
Minkowski space.}
\begin{equation}
x^\mu \rightarrow \tilde{x}{}^\mu=\frac{x^\mu}{x^2},\quad
e\rightarrow \tilde e=\frac{e}{x^2}.
\label{-1}
\end{equation}
The special conformal
transformations (\ref{confin}),
(\ref{efin})
can be treated as
a combination of
(\ref{-1}),
and of the space-time translation for the finite vector
$-\gamma^\mu$
followed by another transformation (\ref{-1}).

In correspondence
with Eqs. (\ref{findil}), (\ref{efin}),
on the phase space
$x^\mu$, $p_\mu$
extended by the canonical pair
$e$ and $p_e$,
the Noether integrals associated with the
scale and special conformal transformations
are corrected,
\begin{equation}
D=xp+2ep_e,\quad K_\mu=2x_{\mu }(xp+2ep_e)-x^{2}p_{\mu }.
\label{corrpe}
\end{equation}
The additional terms, however, vanish on the physical
subspace given
by the constraints
$p_e=0$ and  $p^2=0$,
where (\ref{corrpe}) coincide with the $D$ and $K_\mu$ from
(\ref{genconf}).
Note also that on the physical subspace
the generators $D$ and $K_\mu$
can be represented in terms of the Poincar\'e generators,
$$
D=\frac{1}{P^0}J_{0\nu}P^\nu,\qquad
K_\mu=\frac{1}{(P^0)^2}(2J_{\mu\nu}P_0-
J_{0\nu}P_\mu)J^{0\nu}.
$$

The most noticeable feature
in the structure of the
conformal symmetry
is the nonlinear nature of the
transformations
induced by $K_\mu$.
Since the conformal algebra (\ref{confalg})
is isomorphic to the $so(2,d)$ algebra
(\ref{so24}), and
the latter is the
isometry of AdS${}_{d+1}$,
it is natural to
look at the massless particle
in the context of the AdS/CFT
correspondence.
Before doing this, we shall discuss
briefly the non-relativistic
particle systems possessing the conformal symmetry
$so(2,1)$ in the form of the
dynamical symmetry.

\section{Dynamical $so(2,1)$ Symmetry}

A
free non-relativistic particle
in $\R^d$,
\begin{equation}
L=\frac{m}{2}\dot{x}_i{}^2,\qquad \dot{x}_i=\frac{dx_i}{dt},
\label{mnr}
\end{equation}
possesses the symmetry
\begin{equation}
\delta x^i= \omega^{ij}x^j+\alpha^i+
\upsilon^i t+\beta x^i+\gamma x^i t,\qquad
\delta t=\epsilon + 2\beta t +\gamma t^2.
\label{ncf}
\end{equation}
The transformations given by
the infinitesimal parameters
$\omega^{ij}$,
$\alpha^i$,
$\upsilon^i$ and
$\epsilon$
correspond to the Galilei group symmetry,
whereas
$\beta$ and $\gamma$
are associated with
the scale and special conformal
symmetries. The global form of the
last symmetry transformations is
\[
t'=\frac{t}{1-\gamma t},
\qquad
x'_i=\frac{x_i}{1-\gamma t},
\]
that can be compared with its relativistic
counterpart (\ref{confin}).
The
integrals of motion
corresponding to (\ref{ncf})
are
\begin{equation}
P_i=p_i,\quad
M_{ij}=x_ip_j-x_jp_i,\quad
G_i=tp_i-mx_i,
\label{galilei}
\end{equation}
\begin{equation}
H=\frac{1}{2m} p_i^2,\quad
D=\frac{1}{2}x_ip_i-Ht,\quad
K=\frac{m}{2}x_i^2 -2Dt -Ht^2.
\label{gso}
\end{equation}
The three transformations touch the
time parameter, and their corresponding
generators
(\ref{gso})
form the dynamical symmetry
being the conformal symmetry $so(2,1)$,
\begin{equation}
\{D,H\}=H,\quad
\{D,K\}=-K,\quad
\{H,K\}=-2D.
\label{so21}
\end{equation}
With the identification
$$
J_{01}=\frac{1}{2}(H-K),\quad
J_{02}=\frac{1}{2}(H+K),\quad
J_{12}=D,
$$
the algebra (\ref{so21}) takes the form (\ref{so24})
with $A,B=0,1,2,$  and
$\eta_{AB}=diag (-1,+1,-1)$.
In accordance with Eq. (\ref{gso}),
the $so(2,1)$ quadratic Casimir element is
related to the Casimir element of the rotation
group $SO(d)$,
\begin{equation}
{\cal C}=\frac{1}{2}J_{AB}J^{AB}=-D^2+HK=\frac{1}{8}
M_{ij}^2.
\label{casimir}
\end{equation}

The $so(2,1)$ is also the dynamical symmetry
of the conformal mechanics model of \cite{AFF},
\begin{equation}
L=\frac{m}{2}\dot{x}_i^2-\frac{\alpha}{2x_i^2},
\label{ax}
\end{equation}
given on the configuration space
${\cal R}^d=\R^{d}-\{0\}$.
In particular case of $d=1$
(\ref{ax})
describes the relative
motion in the
2-particle Calogero model \cite{Calog}.
For this system $P_i$ and $G_i$
are
not anymore integrals of motion, but the
$D$ and $K$
of the form (\ref{gso}) with the Hamiltonian
\begin{equation}
H=\frac{1}{2m}p_i^2+\frac{\alpha}{2x_i^2}
\label{h1r}
\end{equation}
are still the integrals of motion which together
form the $so(2,1)$ algebra.
The $so(2,1)$ quadratic Casimir element
is reduced here
to
\begin{equation}
{\cal C}=\frac{1}{4}\left(\frac{1}{2}M_{ij}^2+\alpha
m\right).
\label{calp}
\end{equation}

Another system enjoying
the conformal symmetry is the
non-relativistic particle
in a planar
gravitational field of a point massive source
\cite{Des}. It
has the dynamics of the
free particle
on the cone
described by the Lagrangian
\begin{equation}
L=\frac{m}{2}(\dot{r}{}^2+
\sigma^2 r^2\dot{\varphi}{}^2).
\label{lcone}
\end{equation}
Here we have used the following
parametrization of the cone:
\begin{equation}
x_1=\sigma r\cos\varphi,\quad
x_2=\sigma r\sin\varphi,\quad
x_3=r\sqrt{1-\sigma^{2}},\quad
0<\sigma<1.
\label{conpar}
\end{equation}
In terms of the canonically conjugate
variables
($r$, $p_r$) and ($\varphi$, $J$)
the corresponding $so(2,1)$ generators  are
given by
\begin{equation}
H=\frac{p_r^2}{2m}+\frac{\sigma^{-2}J^2}{2mr^2},\qquad
D=\frac{1}{2}rp_r-Ht,\qquad
K=\frac{m}{2}r^2-2Dt-Ht^2,
\label{confcone}
\end{equation}
and the value of the Casimir element,
${\cal C}=\frac{1}{4}\sigma^{-2}J^2$,
is fixed  by the value of the angular
momentum integral $J$.

The list of the non-relativistic
systems with the dynamical $so(2,1)$ symmetry
includes also the
planar charge-vortex \cite{Jackv} and $3$-dimensional
charge-monopole systems \cite{Jackm,mp}.
The corresponding Lagrangian
\begin{equation}
L=\frac{m}{2}\dot x_i^2+qA_i(x)\dot x_i
\label{vormon}
\end{equation}
describes the particle of charge $q$
either
on the punctured plane
${\cal R}^2$, or in the space
${\cal R}^3$
subjected to the magnetic field
given by  the U(1) gauge potential
$A_i$,
\begin{equation}
\partial_1 A_2-\partial_2 A_1=B=
\frac{\Phi}{2\pi}\delta^{2}(x_i),\quad
x_i\in{\cal R}^2,
\label{avor}
\end{equation}
\begin{equation}
\partial_i A_j-\partial_j A_i=\epsilon_{ijk}B_k,\quad
B_i=g\frac{x_i}{(x_j^2)^{3/2}},\quad
x_i\in {\cal R}^3.
\label{amon}
\end{equation}
The Hamiltonian for the both systems
can be represented as
\begin{equation}
H=\frac{1}{2m}{\cal P}_i^2,
\label{hp}
\end{equation}
with
${\cal P}_i=p_i-qA_i(x)$.
In terms of ${\cal P}_i$,
the angular momentum integral
of the charge-monopole system,
$J_i=\frac{1}{2}\epsilon_{ijk}M_{jk}=\epsilon_{ijk}x_jp_k$,
takes the form
\begin{equation}
J_i=\epsilon_{ijk}x_j{\cal P}_k-
qg n_i,\quad n_i=\frac{x_i}{\sqrt{x_l^2}},
\label{jmono}
\end{equation}
and the angular momentum integral of the
charge-vortex system,
$J=\epsilon_{ij}x_i p_j$,
is
$$
J=\epsilon_{ij}x_i{\cal P}_j
+\frac{q\Phi}{2\pi}.
$$
The Hamiltonian (\ref{hp}), and the $D$ and $K$ given by the
equations
(\ref{gso}) with $p_i$ substituted for
${\cal P}_i$
are the integrals of motion
generating the dynamical
conformal symmetry $so(2,1)$.
Here, the Casimir element is reduced to
$
{\cal C}=\frac{1}{4}(J-\frac{q\Phi}{2\pi})^2
$
for the charge-vortex, and to
$
{\cal C}=\frac{1}{4}(J_i^2-q^2g^2)
$
for the charge-monopole systems.

The  conformal symmetry of the
non-relativistic systems
(\ref{mnr}), (\ref{ax}),
(\ref{lcone})
and (\ref{vormon})
together with the rotation symmetry
makes their dynamics to be very similar.
E.g., as follows from Eq.
(\ref{jmono}),
the trajectory of the particle in the monopole
field is confined to the
cone $\vec J\vec n=-qg$,
and due to the equations
$\ddot{x}_i=-qgr^{-1}\epsilon_{ijk}x_j\dot{x}_k$,
$r=\sqrt{x_l^2}$,
its motion over the cone is free \cite{mp}.
The particle's position
being projected to the plane orthogonal
to $\vec J$ with preservation of the distance
from the origin is
given by the vector
\[
\vec X=r\vec {\cal N},\quad \vec {\cal N}=
\frac{\vec n_\perp}{|\vec n_\perp|},\quad \vec n_\perp=
\vec n-\vec J
\frac{\vec n\vec J}{\vec J^2} ,
\]
which obeys the same equations of motion as the vector
$\vec x$ of the particle in the system
(\ref{ax}) with $\alpha=-q^2 g^2/m<0$,
\[
\ddot{\vec x}=\frac{\alpha}{m}\frac{\vec x}{(\vec x^2)^2}.
\]
The origin of such a similarity
is rooted in the structure
of the corresponding Hamiltonians
which being written in terms of
the canonical radial variables
$r=\sqrt{x_i^2}$ and
$p_r=r^{-1}p_ix_i$ take a general
form
\begin{equation}
H=\frac{p_r^2}{2m}+\frac{{\cal J}^2}{2mr^2},
\label{hamgen}
\end{equation}
where ${\cal J}^2$ is equal to
$\frac{1}{2}M_{ij}^2$ for the
free particle in $\R^d$
(or, in ${\cal R}^d$),
to $\frac{1}{2}M_{ij}^2+\alpha m$
for the model (\ref{ax}),
to $\sigma^{-2} J^2$ for the free particle
on the  cone\footnote{For this case
a slightly different definition (\ref{conpar})
for
the radial variable has been used;
however, the Hamiltonian takes the same form
as in (\ref{confcone})
if we parametrize the cone in terms of the usual
radial variable in the plane,
$r^2=x_1^2+x_2^2$, and then realize
the canonical transformation $r\rightarrow \sigma^{-1} r$,
$p_r\rightarrow \sigma p_r$.},
to $(J-\frac{q\Phi}{2\pi})^2$
for the planar charge-vortex system,
and to $\vec J^2-q^2g^2$ for the charged particle
in the monopole field.
As it was shown recently \cite{Bell},  by a canonical
transformation
the Hamiltonian of the ``new" conformal mechanics
of \cite{kallosh} can be reduced  to that of
the model (\ref{ax}), and so,
(\ref{hamgen}) comprises also
the model \cite{kallosh} (see Section 5).
This general structure will be exploited
below under discussion of the
AdS/CFT correspondence for non-relativistic
systems.

\section{Conformal Symmetry of Massless Particle}
Let us return to the relativistic case.
Having in mind  the identification (\ref{ident}),
we consider a particle
in $(d+2)$-dimensional space
$\R^{2,d}$
with  coordinates
$\X^A$ and metric (\ref{etaab}),
and introduce
canonical momenta
$\P_A$,
$\{\X_A,\P_B\}=\eta_{AB}$.
In terms of $\X_A$ and $\P_A$
the $so(2,d)$ generators
are realized
quadratically,
\begin{equation}
\J_{AB}=\X_A \P_B - \X_B \P_A.
\label{jxp}
\end{equation}
In order the system
would have the same number of degrees
of freedom as the massless scalar
particle in $\R^{1,d-1}$,
we introduce the scalar equations
\begin{eqnarray}
&\phi_0= \P_A\P^A=0,&
\label{xp2}\\
&\phi_1=\X^A\X_A=0,\qquad
\phi_2=\X^A\P_A=0.&\label{xp1}
\end{eqnarray}
The relations $\phi_1=0$
and $\phi_0=0$
could be treated
as the constraints only if
the hypersurfaces
$\X_0=\X_1=\ldots=\X_{d+1}=0$
and $\P_0=\P_1=\ldots=\P_{d+1}=0$
are
excluded from the phase space.
Only in this case the gauge orbits
generated by $\phi_1$ and $\phi_0$
are regular\footnote{
For the massless particle
in $\R^{1,d-1}$,
the points with $p_0=p_1=\ldots=p_{d-1}=0$,
corresponding to the trivial representation
of the Poincar\'e group  have also to be excluded
from the phase space by the
same reasons. This makes a massless particle
to be similar to the non-relativistic
charge-monopole system \cite{mp}.},
and then (\ref{xp2}) and  (\ref{xp1})
form the set of the first class constraints \cite{HT,PR}.
Therefore, the configuration space
of the system (\ref{xp2}), (\ref{xp1})
is supposed to be
 ${\cal R}^{2,d}=\R^{2,d}-\{0\}$.
Note that up to inessential numerical factors
the constraints (\ref{xp2}), (\ref{xp1})
are of the form of the
non-relativistic generators (\ref{gso}) taken
at $t=0$, i.e. they
generate the local $so(2,1)$ symmetry.

As in the $d$-dimensional case,
the $so(2,d)$ generators (\ref{jxp})
can be supplied with
the generators $\P_A$, and
\begin{equation}
\D=\X\P,\qquad
\K_A=2\X_A(\X\P)-\P_A\X^2,
\label{dk}
\end{equation}
which together with
(\ref{jxp}) form the
$so(3,d+1)$ algebra. However,
$\P_A$ do not commute with the
first constraint from (\ref{xp1})
and so,  are not observable.
Since the quantities (\ref{dk})
are proportional to the
constraints (\ref{xp1}),
the global (rigid) symmetries associated
with the generators $\D$ and $\K_A$
are reduced to the local (gauge) symmetries.
On the other hand,
the scalar character of
the constraints (\ref{xp2}),
(\ref{xp1})
means  that the $so(2,d)$
generators (\ref{jxp})
are the observable (gauge-invariant)
quantities and should be related to the
set of the conformal generators (\ref{genconf}).
As we shall see,
such a direct
relation between the symmetry generators
in $d$ and $d+2$ dimensions
is due to the special choice of the constraints (\ref{xp1}),
but generally it can be more involved.

In accordance with the structure
of the constraints (\ref{xp2}), (\ref{xp1}),
the system has a sense of the massless
particle
on the $(d+1)$-dimensional cone
$\X^A\X_A=0$ with local scale symmetry
generated by the constraint $\phi_2$.
The Lagrangian can be chosen, e.g., in one of the
two alternative
forms,
\begin{equation}
L=\frac{\dot\X^2}{2e}+\frac{v}{2}\X^2,
\label{lag+1}
\end{equation}
or
\begin{equation}
L=\frac{(\dot\X-u\X)^2}{2e}+\frac{v}{2}\X^2,
\label{lag+2}
\end{equation}
where $e$, $v$, $u$ are the scalar Lagrange multipliers.
Lagrangian (\ref{lag+2}) produces all the three constraints
(\ref{xp2}), (\ref{xp1}) as the secondary constraints,
whereas for (\ref{lag+1}) the constraint
$\phi_2=0$ appears as the tertiary one.

To establish the exact
relationship between the
$(d+2)$-dimensional massless system
(\ref{xp2}), (\ref{xp1})
for which the $so(2,d)$ is the
Lorentz symmetry, and the $d$-dimensional
massless particle, for which
the same plays a role of the conformal
symmetry, it is convenient
to define
\begin{equation}
\X^{\pm }=\X^{d}\pm \X^{d+1},\quad
\P_{\pm } =\frac{1}{2}\left( \P_{d}\pm
\P_{d+1}\right)
\end{equation}
having the nontrivial Poisson brackets
$\{\X ^{+},\P_{+}\} =
\{\X^{-},\P_{-}\} =1$.
Then,
the $so(2,d)$ generators (\ref{jxp})
take the form
\begin{eqnarray}
&\J_{\mu\nu}=\X_\mu\P_\nu-\X_\nu\P_\mu,\qquad
\J_{d(d+1)} =\X^+\P_+ -\X^-\P_-,&\notag\\
&\J_{\mu+}\equiv \J_{\mu d}+\J_{\mu (d+1)}=
2\X_\mu\P_+ - \P_\mu \X^-,\quad
\J_{\mu-}\equiv\J_{\mu d}-\J_{\mu (d+1)}=
2\X_\mu\P_- - \P_\mu \X^+.&
\label{d+-}
\end{eqnarray}
Generically\footnote{This always can be
achieved by applying the appropriate Lorentz
transformation.} $\X^-\neq 0$, and
having in mind the identification
(\ref{ident}), the symplectic form of the system,
$\Omega=d\P_A\wedge d\X^A=
d\P_\mu\wedge d\X^\mu+d\P_+\wedge d\X^+
+d\P_-\wedge d\X^-$,
can be represented equivalently as
$$
\Omega=d\tilde\P_\mu\wedge d\tilde\X{}^\mu+
d\tilde\P_+\wedge d\tilde \X{}^+
+d\tilde\P_-\wedge d\tilde \X^-,
$$
where
\begin{equation}
\tilde \X^\mu=-\frac{\X^\mu}{\X^-},\quad
\tilde\X^+=\frac{\X_A\X^A}{\X^-},\quad
\tilde\X^-=\X^-,
\label{cantr0}
\end{equation}
\begin{equation}
\tilde \P_\mu=\J_{\mu +},\quad
\tilde\P_+=\P_+,\quad
\tilde\P_-=\frac{1}{\X^{-2}}\left(\X^-\P_A\X^A-\P_+\X_A\X^A
\right).
\label{cantr}
\end{equation}
Therefore, the transformation
$
(\X^A,\P_A)\rightarrow
(\tilde\X{}^A,\tilde\P_A)
$
is  canonical, and
its inverse is
$$
\X_\mu=-\tilde\X^-\tilde\X_\mu,\quad
\X^-=\tilde\X^-,\quad
\X^+=\tilde\X^+-\tilde\X^-\tilde\X_\mu\tilde\X^\mu,
$$
$$
\P_\mu=-\frac{\tilde\P_\mu}{\tilde\X^-}-2\P_+\tilde\X_\mu,
\quad
\P_+=\tilde\P_+,\quad
\P_-=\tilde\P_--\tilde\P_+\tilde\X_\mu\tilde\X^\mu-
\frac{\tilde\X_\mu\tilde\P^\mu}{\tilde\X^-}.
$$
In terms of the new variables the constraints
read as
\begin{equation}
\phi_0=\tilde\P_\mu\tilde\P^\mu+
4\frac{\tilde\P_+}{\tilde\X^{-2}}(\tilde\X^-\phi_2-
\tilde\P_+\phi_1)=0,
\label{tphi0}
\end{equation}
\begin{equation}
\phi_1=\tilde\X^+\tilde\X^-=0,\qquad
\phi_2=\tilde\X^+\tilde\P_++\tilde\X^-\tilde\P_-=0,
\label{contil}
\end{equation}
and the Lorentz generators (\ref{d+-})
take the form
$$
\J_{\mu\nu}=\tilde\X_\mu\tilde\P_\nu-
\tilde\X_\nu\tilde\P_\mu,\quad
\J_{\mu +}=\tilde\P_\mu,\quad
\J_{d(d+1)}=\tilde\X_\mu\tilde\P^\mu+
2\frac{\tilde\P_+}{\tilde\X^-}\phi_1
-\phi_2,
$$
$$
\J_{\mu -}=2(\tilde\X_\nu\tilde\P^\nu)\tilde\X_\mu
-(\tilde\X_\nu\tilde\X^\nu)\tilde\P_\mu
+\frac{1}{\tilde\X^{-2}}(\tilde\P_\mu
+4\tilde\P_+\tilde\X^-\tilde\X_\mu)\phi_1
-2\tilde\X_\mu\phi_2.
$$
The constraints
(\ref{contil})
single out the surface
\begin{eqnarray}
&\tilde\P_\mu\tilde\P^\mu=0,&\label{tp2}
\\
&\tilde\X^+=0,\qquad
\tilde\P_-=0.&\label{x+p-}
\end{eqnarray}
Therefore, the variables $\tilde\P_\mu$ and
$\tilde\X_\mu$ are the observable (gauge invariant)
variables with respect to the constraints
(\ref{xp1}), whereas
$\tilde\X^-=\X^-$ and
$\tilde\P_+=\P_+$
are the pure gauge variables which
can be removed by introducing
the constraints
\begin{equation}
\phi_3\equiv\X^-+1=0,\quad
\phi_4\equiv\P_+=0
\label{gaugex}
\end{equation}
as the gauge conditions for (\ref{xp1}).
Reducing the system to the surface given by
the set of the second class constraints (\ref{xp1}),
(\ref{gaugex})
 being equivalent to the set (\ref{x+p-})
(\ref{gaugex}),
we completely exclude from the theory
the variables $\X^\pm$, $\P_\pm$,
and
as a result,
the canonical variables $\tilde\X^\mu$ and $\tilde\P_\mu$
are reduced to the initial variables
$\X^\mu$ and $\P_\mu$,
the mass shell constraint takes
the form $\phi_0=\P_\mu\P^\mu=0$,
and the Lorentz generators (\ref{jxp})
are reduced to the $d$-dimensional generators
(\ref{genconf}) of the conformal symmetry.

So, the
initial $d$-dimensional massless system (\ref{m0})
can be reinterpreted as the
$(d+2)$-dimensional massless system
(\ref{lag+1}) or (\ref{lag+2}) living on the cone
and possessing
local scale symmetry.
Under such identification
the $so(2,d)$ Lorentz generators
of the $(d+2)$-dimensional system
correspond to the conformal symmetry
generators in accordance with identification
(\ref{ident}).
In this way, the cubic structure of the
special conformal symmetry generators
$K_\mu$ originates from the nonlinearity
of the canonical transformation (\ref{cantr0}),
(\ref{cantr}).

The constraints (\ref{gaugex}) and (\ref{xp2})
forming  the set of the first class constraints
can be used instead of the set
(\ref{xp2}), (\ref{xp1}).
Having in mind the origin of the relations
(\ref{x+p-}), the change of the constraints
(\ref{xp1}) for
the constraints
(\ref{gaugex}) effectively is a canonical transformation
acting nontrivially on the Lorentz generators (\ref{jxp}).
We shall not investigate such an action explicitly,
but,
instead, look at the rigid symmetries
for the case of the choice of the first class constraints
(\ref{gaugex}).
The variables $\X_\mu$ and $\P_\mu$
are the observables and can be identified with
$x_\mu$ and $p_\mu$ of the $d$-dimensional model.
Then, on the surface of the constraints (\ref{gaugex})
in addition to the Poincar\'e generators
$J_{\mu\nu}=\J_{\mu\nu}$, $P_\mu=\P_\mu=\J_{\mu +}$,
the scale and the special conformal
generators from (\ref{genconf}) can be identified
with the following linear combinations
of the $so(3,d+1)$ generators:
$D=\P_-+\D=P_--(\K_d+\K_{d+1})$,
$K_\mu=\K_\mu+\J_{\mu-}$,
which are the gauge invariant quantities.
So, when the system is given by
the constraints (\ref{xp2}),
(\ref{gaugex}),
the identification of the
generators $D$ and $K_\mu$ includes
the $so(3,d+1)$ generators
$\D$ and $\K_A$.
With such a treatment the nonlinearity of the special
conformal symmetry transformations
is rooted in the nonquadratic nature of $\K_A$
given by Eq. (\ref{dk}).

The $(d+2)$-dimensional system with the
first class constraints
(\ref{xp2}), (\ref{gaugex})
can be described by the  Lagrangian
\begin{equation}
L=\frac{\dot{\X}_A\dot{\X}{}^A}{2e}
-v(\X^-+1).
\label{lag+x}
\end{equation}
Due to the direct identification of the
observables of the $(d+2)$-dimensional
system (\ref{lag+x})
with the phase space coordinates of the
$d$-dimensional system, the corresponding
field formulations for the both systems
are also related in a very simple way.
In accordance with Eqs. (\ref{gaugex})
and (\ref{tp2}),
the field $\Phi(\X_A)$ satisfies
the equations
$$
(\X^-+1)\Phi(\X_A)=0,\qquad
\partial_+\Phi(\X_A)=0,\qquad
\partial_B^2\Phi(\X_A)=0,
$$
where
$\partial_B=\partial/\partial \X^B$,
$\partial_+=\partial/\partial \X^+$.
Their solution is
$\Phi(\X_A)=
\delta(X^-+1)\varphi(\X_\nu)$
with the field
$\varphi(\X_\nu)$
obeying the $d$-dimensional Klein-Gordon
equation $\partial_\mu^2\varphi(\X_\nu)=0$,
$\partial_\mu=\partial/\partial \X^\mu$.

The $(d+2)$-dimensional system
(\ref{xp2}), (\ref{xp1})
can be reinterpreted
as a massless particle living on
the
border of the AdS${}_{d+1}$ of infinite radius,
whose isometry corresponds to the
conformal symmetry of the massless particle
in $d$ dimensions.
This can be done in the following way.
The massless particle
on AdS${}_{d+1}$ of radius $R$
can be described by the
constraints
\begin{equation}
\underline{\phi}_0=\underline{\P}{}_A
\underline{\P}{}^A=0,\qquad
\underline{\phi}_1=\underline{\X}{}_A\underline{\X}^A+R^2=0,
\label{adsl}
\end{equation}
where $\underline{\X}^A$ and
$\underline{\P}{}_A$ are the canonical variables.
The Poisson brackets of
the constraints (\ref{adsl})
are $\{\underline{\phi}_1,\underline{\phi}_0\}=
4\underline{\X}^A\underline{\P}{}_A$.
To have the reparametrization-invariant system,
the $\underline{ \phi}_0$ has to be
the first class constraint.
This can be achieved by
postulating the constraint
\begin{equation}
\underline{\phi}_2=\underline{\X}^A\underline{\P}{}_A=0
\label{scale}
\end{equation}
in addition to the constraints (\ref{adsl}).
The constraint (\ref{scale}) generates
the local scale transformations
which due to the relation $\{\underline{\phi}_2,
\underline{\phi}_0\}=2\underline{\phi}_0$
are consistent with the
reparametrization invariance
generated by the constraint $\underline{\phi}_0$.
Since
$\{\underline{\phi}_2,
\underline{\phi}_1\}=-\underline{\phi}_1+R^2\neq0$,
the constraint $\underline{\phi}_1=0$ can be understood
as a gauge condition for the constraint
(\ref{scale}).
Now, let us realize a canonical transformation
$(\underline{\X}^A,\underline{\P}_A)
\rightarrow
(\X^A,\P_A)$,
\begin{equation}
\X^A=\frac{\underline{\X}^A}{R^{1+\varepsilon}},
\quad
\P_A=\underline{\P}{}_AR^{1+\varepsilon},
\label{trick}
\end{equation}
with a constant $\varepsilon>0$,
and take a limit $R\rightarrow\infty$,
$\underline{\X}^A\rightarrow\infty$,
$\underline{\P}{}_A\rightarrow 0$
in such a way that the variables
(\ref{trick}) would be finite.
Then, the constraints
(\ref{adsl}), (\ref{scale}) take the form
of the first class constraints (\ref{xp2}), (\ref{xp1}),
and we reproduce the
system on the cone.
Because of the change of the nature
of the constraints from the second
to the first class,
the described limit procedure
has a rather formal character;
however, see the next Section for further discussion of
the system (\ref{adsl}), (\ref{scale}).

\section{AdS/CFT Correspondence for Non-Relativistic
Systems}

Now, let us show that the dynamical $so(2,1)$
symmetry of the non-relativistic systems discussed in
Section 3
can be naturally understood in the context of the
CFT/AdS correspondence.
We first analyze in detail the simplest
system of the free massive particle
(\ref{mnr}), and then generalize the results
for other non-relativistic systems.

Let us consider the massive particle
(\ref{mnr}) in
$\R^1$. Changing the parametrization
of the trajectory,
$x^1(t)\rightarrow x^1(t(\tau))$,
$\dot{t}(\tau)=\frac{dt}{d\tau}>0$,
and introducing the notation $t=x^0$,
we pass from (\ref{mnr}) to the
reparametrization-invariant action
\begin{equation}
A=\int Ld\tau,\qquad
L=\frac{m}{2}\frac{\dot{x}_\mu^2}{\dot{x}{}^0}.
\label{l2}
\end{equation}
Here $\dot{x}_\mu^2=\dot x{}^\mu\dot x{}^\nu\eta_{\mu\nu}$,
$\eta_{\mu\nu}=diag (-1,1)$,
and we have added a
total derivative term
$-\frac{m}{2}\dot x{}^0$.
A simple comparison of (\ref{l2})
being the reparametrization-invariant representation
of the {\it non-relativistic massive} particle
on the one hand
with the {\it massless} particle (\ref{m0})
in the $2$-dimensional
Minkowski space $\R^{1,1}$ on the other hand
reveals a
similarity of the both systems, and  in what follows we
shall
demonstrate that they are canonically related.
Before turning to the Hamiltonian description
of the systems, it is instructive to compare them
in the Lagrangian picture.

The general solution to the equations of motion for
(\ref{l2}),
\begin{equation}
\frac{d}{d\tau}
\left(
\frac{\dot{x}{}^1}{\dot{x}{}^0}\right)=0,
\qquad
\frac{d}{d\tau}
\left(
\frac{\dot{x}{}^1}{\dot{x}{}^0}\right)^2=0,
\label{x1x0}
\end{equation}
can be represented in the form
$x^1=x^1(x^0)$,
\begin{equation}
x^1(x^0)={\rm v} x^0+a,
\label{xva}
\end{equation}
where $a$ and ${\rm v}$ are the integration constants.
This can be compared with the general solution,
\begin{equation}
x^1(x^0)=\epsilon x^0+b,\qquad
e=M^{-1} \dot{x}{}^0,
\label{xeb}
\end{equation}
to the equations of motion
\[
(\dot{x}{}^{1})^2-(\dot{x}{}^0)^2=0,\qquad
\frac{d}{d\tau}
\left(
\frac{\dot{x}{}^1}{e}\right)=0,\qquad
\frac{d}{d\tau}
\left(
\frac{\dot{x}{}^0}{e}\right)=0
\]
for the massless particle,
where $\epsilon=\pm $, and $b$ and  $M$ are the integration
constants
with $M$ having a dimension of mass.
The only formal difference
between (\ref{xva}) and (\ref{xeb})
is in the velocity
of the particles: for the first system it can be arbitrary,
whereas for the second it is the velocity of light.
However, at the moment we note that
for ${\rm v}\neq 0$
the solution (\ref{xva}) can be
reduced to (\ref{xeb})
by rescaling $x^0$.

One can construct the Lagrangian
relating the two systems,
\begin{equation}
L=\frac{\dot{x}_\mu^{2}}{2e}
+\frac{\lambda}{l}(e-M^{-1}\dot{x}{}^0),
\label{inter}
\end{equation}
where $\lambda$ is a Lagrange multiplier
and $l$ is a numerical parameter.
The equations of motion for the  Lagrange multipliers
$\lambda$ and $e$
are algebraic,
\begin{equation}
e-M^{-1}\dot{x}{}^0=0,\quad
(\dot{x}{}^1)^2-(\dot{x}{}^0)^2-\frac{\lambda}{l}e^2=0,
\label{elam}
\end{equation}
the equation for $x^1$
is of the same form as for the massless particle,
and for $x^0$ it is
\[
\frac{d}{d\tau}
\left(
\dot{x}{}^0+\frac{\lambda}{2Ml}\right)=0.
\]
The equations (\ref{elam})
can be solved to represent the
Lagrange multipliers in terms of other
variables,
$
e=M^{-1}\dot{x}{}^0,
$
$
\lambda=M^2 l(({\dot{x}{}^1}/\dot{x}{}^0)^2-1).
$
As a result,
the equations
for $x^1$ and $x^0$
are reduced exactly
to the equations (\ref{x1x0}) for the first system,
that proves the equivalence of the system
(\ref{inter}) to the system (\ref{l2})
(the substitution of the solution for $e$ and $\lambda$
into (\ref{inter}) with identification $M=m$
reduces the latter to (\ref{l2})).
On the other hand,
in the limit $l\rightarrow \infty$
the system (\ref{m0}) is
reproduced from (\ref{inter}).

Now, let us compare the two systems within the
Hamiltonian picture.
In terms of the
light-cone variables $x^\pm=x^0\pm x^1$,
 $p_\pm=\frac{1}{2}(p_0\pm p_1)$,
 $\{x^+,p_+\}=\{x^-,p_-\}=1$,
the $so(2,2)$
generators (\ref{genconf})
of the massless particle in $\R^{1,1}$
are represented in the form
\begin{eqnarray}
& \D_+=\frac{1}{2}(D+M^{01})=x^+p_+,\quad
\mH_+=\frac{1}{2}( P_0+P_1)=p_+,\quad
 \K_+=\frac{1}{2}(K^0+K^1)=x^+{}^2p_+,
&\nonumber\\
&&\label{gen+-}
\\
&
\D_-= \frac{1}{2}(D-M^{01})=x^-p_-,\quad
\mH_-=\frac{1}{2} (P_0-P_1)=p_-,\quad
\K_-=\frac{1}{2}(K^0-K^1)=x^-{}^2p_-.
&
\nonumber
\end{eqnarray}
The subsets
$\D_+,$ $\mH_+$, $\K_+$
and $\D_-,$ $\mH_-$, $\K_-$
mutually commute and every
of them generates
the algebra of the form (\ref{so21}), that
reveals explicitly the direct sum structure of the conformal
symmetry algebra
 $so(2,2)=so(2,1)\oplus so(2,1)$.
Since the constraint
$p^2=-\frac{1}{4}p_+p_-=0$ is equivalent to either
$p_+=0$ or $p_-=0$, on the mass shell
one of the two sets of the $so(2,1)$ generators
turns into zero, and
the corresponding
rigid $so(2,1)$ symmetry is reduced to
the gauge (reparametrization) symmetry,
that leaves only one rigid $so(2,1)$ symmetry.

The non-relativistic system
(\ref{l2}) is characterized by the
constraint
\begin{equation}
\varphi=p_0+\frac{p_1^2}{2m}=0.
\label{cp0}
\end{equation}
Here we have shifted the momentum $p_0$
canonically conjugated to $x^0$,
$p_0+\frac{m}{2}\rightarrow p_0$.
This change is a canonical transformation
corresponding to omitting
the
total derivative term $-\frac{m}{2}\dot{x}{}^0$
from Lagrangian (\ref{l2}).
The quantities
\begin{equation}
T=p^0,\qquad D=\frac{1}{2}p_1x^1+p_0x^0,\qquad
K=\frac{m}{2}x_1^2-p_0(x^0)^2-x^0p_1x^1
\label{integ}
\end{equation}
are the integrals of motion being the generators
of conformal symmetry $so(2,1)$.
The quantization of the system
transforms the constraint (\ref{cp0})
into the Schr\"odinger
equation
\[
i\frac{\partial}{\partial x^0}\psi(x^0,x^1)=
-\frac{1}{2m}\frac{\partial^2}{\partial x_1^2}
\psi(x^0,x^1).
\]
In
such a way the hidden conformal symmetry of the
non-relativistic Schr\"odinger
equation \cite{BaRa}
obtains a
natural explanation as the symmetry
generated by the quantum
analogues of the usual integrals of motion
(\ref{integ})
of the non-relativistic free particle model
represented in a reparametrization invariant form.

The  equivalent representation of (\ref{integ}),
$$
T=H-\varphi,\quad
D=\frac{1}{2}x^1p_1-Hx^0+x^0\varphi,\quad
K=\frac{m}{2}x_1^2-2Dx^0-H(x^0)^2+(x^0)^2\varphi
$$
with $H=\frac{p_1^2}{2m}$,
shows that the
$so(2,1)$ generators in the
reparametrization
invariant
and usual formulations (see Eqs. (\ref{gso}))
coincide up to the terms proportional to the constraint,
and, consequently,
the transformations generated by them
are the same up to the reparametrization.

The two other integrals of motion correspond to the
translation
and the Galilei boost symmetries,
\begin{equation}
P_1=p_1,\quad
X_1=x_1-\frac{x^0}{m}p_1,
\label{galican}
\end{equation}
(cf. Eq. (\ref{galilei})),
from which another set of the
$so(2,1)$ generators
can be constructed,
\begin{equation}
\bar T=\frac{P_1^2}{2m},\qquad
\bar D=\frac{1}{2}X_1P_1,\qquad
\bar K=\frac{m}{2}X_1^2.
\label{intbar}
\end{equation}
However, there is only one independent set
of the $so(2,1)$ generators since the linear
combination of
(\ref{integ}) and (\ref{intbar})
vanishes on the physical subspace given
by the constraint (\ref{cp0}),
$$
T-\bar T=\varphi,\quad
D-\bar D=x^0\varphi,\quad
K-\bar K=(x^0)^2\varphi.
$$

The integrals of motion (\ref{galican})
together  with the relations
\begin{equation}
X^0=x^0,\qquad P_0=p_0+\frac{p_1^2}{2m},
\label{x0p0}
\end{equation}
can be treated as the change of the variables
$x^\mu,p_\mu\rightarrow X^\mu, P_\mu$,
which is a canonical transformation.
Since the $P_0$ coincides with the constraint,
the canonical pair $X^0$ and $P_0$ corresponds to the gauge
degree of freedom, while
$X_1$ and $P_1$ form a pair of the gauge-invariant
independent observables.

Supposing $p_1\neq 0$,
one can consider another canonical transformation
\begin{equation}
\tilde x_1=\epsilon m\frac{x_1}{p_1}, \qquad
\tilde p_1=\epsilon \frac{p_1^2}{2m},\qquad
\tilde x_0=x_0,\qquad
\tilde p_0=p_0,
\label{canon}
\end{equation}
where
$\epsilon={p_1}/{|p_1|}$.
Then the constraint (\ref{cp0}) takes a form
$\varphi=\tilde p_0+\tilde p_1=0$ for $\tilde p_1>0$
and $\varphi=\tilde p_0-\tilde p_1=0$ for $\tilde p_1<0$.
In these variables,
the $so(2,1)$ generators
(\ref{integ}) are
\[
T=\tilde p{}^0,\quad
D=\tilde x^\mu \tilde p_\mu,
\quad
K=2\tilde x_0 (\tilde x_\mu \tilde p^\mu)
-\tilde x{}^2_\mu\tilde p_0 +\tilde x{}^2_\mu\varphi,
\]
that can be compared with the
generators
$P^0$,
$D$ and $K_0$ from
(\ref{genconf})
for the massless particle.
On the other hand,
the generators (\ref{intbar})
being rewritten in the variables
(\ref{canon}),
\[
\bar T=\epsilon p_1,\quad
\bar D= \epsilon(\tilde x_0\tilde p_1-\tilde x_1\tilde p_0
+\tilde x_1\varphi),
\quad
\bar K=\epsilon(2\tilde x_1\tilde x{}^\mu\tilde p_\mu
-\tilde p_1\tilde x_\mu^2+2\tilde x_0\tilde x_1\varphi),
\]
can be compared with the generators
$P_1$, $M_{01}$ and $K_1$
from (\ref{genconf}).
Note that the $\bar D$, being the
generator of the scale
transformations of $X_1$ and $P_1$,
corresponds
to the Lorentz boost generator $M_{01}$,
while the generator of the
special conformal transformations,
$\bar K$, being
quadratic in the generator of the Galilei boosts,
is mapped by (\ref{canon})
into the special conformal symmetry generator
$K_1$  having a cubic structure in
the massless particle's canonical variables.

In accordance with the constraint
(\ref{cp0}),
the physical states
of the non-relativistic particle
with $p_1\neq 0$  are in the $p_0<0$ sector
of the phase space.
The change
$x^0\rightarrow -x^0$ in Lagrangian (\ref{l2})
results in the physical subspace with $p_0>0$.
This corresponds to the two disjoint sectors with
$p^0>0$
and $p^0<0$
in the case of the massless particle.
So, generally the physical states
of the non-relativistic massive particle
obeying the constraint (\ref{cp0})
can be separated into the three sectors:
with $p_0<0$ or $p_0>0$ for which $p_1\neq 0$,
and with $p_0=p_1=0$.
In the last case the
evolution degenerates,
$x_1=a=const$.
For the massless particle,
the analogous sector
with $p_0=p_1=0$
is excluded from the consideration
as corresponding to the trivial representation
of the $(1+1)$-dimensional Poincar\'e group.
The two nontrivial sectors
with $p_0\neq 0$
can be  characterized
by the same rigid $so(2,1)$ symmetry, and
there, as we have seen,
the two systems, massless
and non-relativistic massive,
can be related via the canonical transformation
(\ref{canon}).

The massless particle on AdS${}_2$
of radius $R$
with metric \cite{Ahar}
$$
ds^2=R^2\frac{d z^2-dx_0^2}{z^2}
$$
is described by the Lagrangian
\begin{equation}
L=R^2\frac{\dot z^2-\dot x_0^2}{2ez^2},
\label{m0ads}
\end{equation}
where $z>0$ or $z<0$.
Since Lagrangian (\ref{m0ads}) is reduced to
(\ref{m0}) by a simple change of the
Lagrange multiplier,
$ez^2R^{-2}\rightarrow e$,
and identification of $z$ with $x^1$,
the massive non-relativistic particle
on ${\cal R}^1$ can be treated  as
the system
canonically equivalent to
the massless particle on AdS${}_2$.
Due to exclusion of
$x^1=0$ from the configuration space,
in this case there is neither translation nor
Galilei boost invariance in the non-relativistic
system,
whereas the remaining conformal $so(2,1)$
symmetry corresponds to the isometry of
the configuration space
AdS${}_2$.

Another form of the Lagrangian for the massless particle
on AdS${}_2$ can be chosen
similarly to (\ref{lag+1}),
\begin{equation}
L=\frac{\dot{\X}{}^2}{2e}+
\frac{v}{2}(\X^2 +R^2),
\label{ads3con}
\end{equation}
where
$\X^2=\X^A\X^B\eta_{AB}=
-\X_0^2+\X_1^2-\X_2^2$.
Lagrangian (\ref{ads3con})
generates the constraints
\begin{eqnarray}
&\phi_0=\P^2=0,&\label{phi0}\\
&\phi_1=\X^2+R^2=0,\qquad
\phi_2=\X\P=0,&
\label{phi12}
\end{eqnarray}
and so, corresponds to the particular $d=1$ case
of the massless system (\ref{adsl}), (\ref{scale}).
Due to the scalar nature of the constraints
(\ref{phi0}), (\ref{phi12}),
the $so(2,1)$ generators
$\J_{AB}=\X_A\P_B-\X_B\P_A$
are the integrals of motion being
the isometry generators of AdS${}_2$.
The mass shell constraint (\ref{phi0})
is the first class constraint
generating the reparametrization symmetry,
whereas the constraints (\ref{phi12})
form the subsystem of the second class constraints.
With taking into account
(\ref{phi12}), the mass shell constraint
is represented equivalently in the form
$\J_{AB}\J^{AB}=0$ corresponding to Eq. (\ref{casimir})
for the non-relativistic particle
in ${\cal R}^1$.

The surface of the second class constraints
(\ref{phi12})
can be parametrized as
\begin{equation}
\X^0=R\frac{x^0}{x^1},\quad
\X^1-\X^2=-\frac{R^2}{x^1},\quad
\X^1+\X^2=-\frac{x^+x^-}{x^1},
\label{adspara}
\end{equation}
\begin{equation}
\P_0=\frac{p_+x^+-p_-x^-}{R},\quad
\P_1+\P_2=p_+-p_-,\quad
\P_1-\P_2=\frac{p_+x^{+2}-p_-x^{-2}}{R^2},
\label{adspar}
\end{equation}
in terms of the phase space
variables
$x^\pm=x^0\pm x^1$,
$p_\pm=\frac{1}{2}(p_0\pm p_1)$,
with $x^1>0$ or $x^1<0$.
The parametrization (\ref{adspar}) reduces
the symplectic two-form
$d\P_a\wedge d\X^a$ to
$dp_+\wedge dx^++dp_-\wedge dx^-$,
i.e. the variables $x^+$, $p_+$ and $x^-$, $p_-$
form the two pairs of canonically conjugate variables
on the reduced phase space given by the second
class constraints (\ref{phi12}).
On the surface (\ref{phi12})
the integrals of motion $\J_{AB}$
are reduced to
\begin{equation}
\J_{12}=D,\quad
\J_{01}+\J_{02}=-RT,\quad
\J_{01}-\J_{02}=-\frac{1}{R}K,
\label{direct}
\end{equation}
with $D=p_+x^++p_-x^-$,
$T=p_++p_-$, $K=
p_+x^{+2}+p_-x^{-2}$.
Up to inessential
numerical factors
the $so(2,1)$ isometry
generators (\ref{direct})
have the form of the sum of the corresponding
generators (\ref{gen+-}).
The parametrization (\ref{adspara})
transforms the Lagrangian (\ref{ads3con})
into (\ref{m0ads}) with identification $z=x^1$,
i.e. (\ref{m0ads})
gives the dynamics of the system on the
reduced phase space, where
the first class constraint (\ref{phi0})
is represented as
\begin{equation}
\phi_0=-\frac{1}{2R^2}\J_{AB}\J^{AB}=4x_1^2p_+p_-=0
\label{phi00}
\end{equation}
Another alternative form of the
Lagrangian for the massless particle on AdS${}_2$ is
\begin{equation}
L=-R^2\frac{\dot{\hat \X}{}^2}{2e},
\label{n2}
\end{equation}
where we  assume that
$\X^2<0$,
and $\hat \X_A=\frac{\X_A}{\sqrt{-\X^2}}$ is a unit vector,
$\hat \X^2=-1$.
Lagrangian (\ref{n2})
generates the two
first class constraints
$\phi_0=0$ and $\phi_2=0$
given by Eqs. (\ref{phi0}), (\ref{phi12}).
The Lagrangian (\ref{n2})
in addition to reparametrization invariance
has also local scale invariance.
The latter gauge invariance
can be fixed by introducing
the constraint $\phi_1=\X^2+R^2=0$
as a gauge condition to the constraint
$\phi_2=0$, that means the
equivalence of the systems given by
Eqs. (\ref{ads3con})
and (\ref{n2}).
Therefore, Lagrangians
(\ref{m0ads}), (\ref{ads3con}),
and (\ref{n2})
describe the same system of the
massless particle on AdS${}_2$
being canonically equivalent
to the massless particle in ${\cal R}^{1,1}$.
The latter, as we have seen,
can be related to the free non-relativistic massive particle
in ${\cal R}^1$
via the construction of the reparametrization invariant
form of the action and
canonical transformation (\ref{canon}).

The case of the relativistic massive scalar particle
on AdS${}_2$ is obtained by adding
the ``cosmological" term $-\frac{e}{2}{\cal M}^2$
to the Lagrangians (\ref{m0ads}),
(\ref{ads3con}), or (\ref{n2}).
This changes the mass shell constraint
(\ref{phi0}) for $\phi_M=\P^2+{\cal M}^2=0$, and its reduced
phase
space  form
(\ref{phi00}) for
\begin{equation}
\phi_M=4x_1^2p_+p_--{\cal M}^2=0.
\label{phi0m}
\end{equation}
The constraint (\ref{phi0m}) is equivalent to the
relation $\frac{1}{2}\J_{AB}\J^{AB}=R^2{\cal M}^2$,
the comparison of which with (\ref{calp}) in the case
of 1-dimensional space ($M_{ij}=0$)
reveals the similarity of the relativistic
massive particle system on AdS${}_2$ with
the non-relativistic conformally invariant
system (\ref{ax}) with the
$m\alpha$ identified with ${\cal M}^2R^2$.
The relation between the two systems
can be established explicitly
via identification of  $z$ with
$x_1$ and
redefinition of the Lagrange multiplier,
$ezR^{-2}\rightarrow e$,
that presents the Lagrangian (\ref{m0ads}) with added
``cosmological" term in the form
\begin{equation}
L=\frac{\dot{x}_1^2-\dot{x}_0^2}{2e}
-\frac{e}{2}\frac{{\cal M}^2R^2}{x_1^2}.
\label{lagmasads}
\end{equation}
This Lagrangian can be compared with
\begin{equation}
L=m\frac{\dot{x}_\mu^2}{2\dot{x}{}^0}
-\frac{\dot{x}{}^0}{2}\frac{\alpha}{x_1^2}
\label{repar2}
\end{equation}
being the reparametrization invariant version
of the Lagrangian (\ref{ax})
for the non-relativistic particle in ${\cal R}^1$.
With the parameter identification
$m\alpha={\cal M}^2R^2$,
the canonical equivalence of the non-relativistic
system (\ref{ax}) in ${\cal R}^1$ to the
massive relativistic particle on AdS${}_2$
can be established in the same way as it was done for the
systems of
the free non-relativistic and massless
particles (see also \cite{Bell}).

The generalization of (\ref{n2}) for the case of
arbitrary dimension
is given
by the Lagrangian describing a relativistic particle
on the AdS${}_2\times$S${}^{d-1}$:
\begin{equation}
L=R^2\frac{\sigma^2 \dot{\hat{\vec N}}{}^2-\dot{\hat \X}{}^
2}{2e}
-\frac{e}{2}{\cal M}^2+{\mathfrak T}_d(\hat {\vec N}).
\label{adss}
\end{equation}
Here $\hat N_i=N_i/\sqrt{\vec N^2}$,
$\vec N\in {\cal R}^d$,
is a unit vector,
$\hat{\vec N}{}^2=1$, and
${\mathfrak T}_d(\hat {\vec N})$ is a topologically
nontrivial term
whose $d=2$ and $d=3$ form
is specified below.
The system (\ref{adss}) with ${\cal M}=0$,
${\mathfrak T}_d(\hat {\vec N})=0$,
and $\sigma=1$ may be canonically related
to the free non-relativistic
particle in ${\cal R}^d$,
while the choice $0<\sigma<1$
corresponds to
the $d$-dimensional generalization
of the free non-relativistic particle
on the cone (\ref{lcone}) obtained via the
substitution
$\sigma^2 r^2\dot{\varphi}^2\rightarrow \sigma^2
r^2\dot{\vec n}{}^2$, $\vec n\in S^{d-1}$.
The case with $\sigma=1$, ${\mathfrak T}_d(\hat {\vec N})=0$
and ${\cal M}\neq 0$ corresponds to the system
(\ref{ax}) with $\alpha>0$,
whereas the case $\alpha<0$ of the model (\ref{ax})
is related to (\ref{adss}) with
${\cal M}^2$ changed for $-{\cal M}^2$.
Such a modified system (\ref{adss})
describes the free tachyon particle on
the AdS${}_2\times $S${}^{d-1}$.
At last, the charge-vortex and the
charge-monopole systems
correspond to the choice $\sigma=1$,
${\cal M}=0$ and
\cite{Jackv,mp,PR}\footnote{
Up to a numerical coefficient,
the term ${\mathfrak T}_d(\hat {\vec N})$
has a sense of the highest curvature (torsion)
of some smooth curve in $\R^d$ to which
the vector $\hat{\vec N}(\tau)$ is tangent
with parameter $\tau$ treated as the natural
parameter \cite{Plt}.}
\[
{\mathfrak T}_2(\hat {\vec N})=
\frac{q\Phi}{2\pi}\epsilon_{ij}\hat N_i\dot{\hat N}_j,\qquad
{\mathfrak T}_3(\hat {\vec N})=-\frac{qg}{\dot{\vec{\hat
N}}{}^2}
\epsilon_{ijk}\hat N_i
\dot{\hat N}_j\ddot{\hat N}_k.
\]
To demonstrate that
the Lagrangian (\ref{adss}), corresponds, e.g.,
in the last case to
the specified non-relativistic systems,
we write the
reparametrization invariant version
for the Lagrangian (\ref{vormon})
with the gauge potential given by Eqs.
(\ref{avor}),
and (\ref{amon}):
\begin{equation}
L=m\frac{r^2\dot{\vec n}^2+
\dot{r}^2-\dot{x}_0^2}{2\dot{x}{}^0}+
{\mathfrak T}_d(\vec n).
\label{tphi}
\end{equation}
Here we have presented $x_i\in {\cal R}^d$,
$d=2,3$,
as $x_i=rn_i$,
$\vec n^2=1$.
Parametrizing the
unit vector $\vec n$ by the polar,
$\vec n(\varphi)=(\cos \varphi,\sin \varphi)$,
and spherical,
$\vec n(\varphi, \vartheta)=(\sin\vartheta\cos\varphi,
\sin\vartheta\sin\varphi,\cos\vartheta)$,
(local) angle variables,
we get the following representation for
the interaction term $qA_i\dot x_i$:
${\mathfrak T}_2=\frac{q\Phi}{2\pi}\dot{\varphi}$
and  ${\mathfrak T}_3=qg\cos \vartheta \dot{\varphi}$
for the planar charge-vortex
and the
$3D$ charge-monopole systems, respectively.
Returning to
the Lagrangian (\ref{adss}),
we note that in addition to
reparametrization it possesses the
two local scale symmetries
acting independently on the variables $\X_A$ and $N_i$.
The scale gauge invariance can be fixed
directly in the Lagrangian
by putting in it $\X^2=-R^2$ and $\vec N^2=1$.
Then,  parametrizing $\X_A$ as in
(\ref{adspara}) and $N_i$ in terms of
the polar or spherical coordinates,
and identifying
$x^1$ from (\ref{adspar})
with the radial variable $r$,
we establish the canonical relation
between (\ref{adss})
and (\ref{tphi}) in the same way
as it was done for the free non-relativistic
and massless particles:
by redefinition of the Lagrange multiplier
and  application of canonical transformation (\ref{canon}).

The general case of the
$so(2,1)\oplus so(d-1)$ invariant Lagrangian (\ref{adss})
corresponds to the
superposition
of the non-relativistic systems considered in Section 3.
E.g., the $d=2$ case with $0<\sigma<1$, ${\cal M}\neq 0$
and nontrivial ${\mathfrak T}_2$
describes the charged particle on the cone
in the presence of the
scalar potential $1/r^2$ and magnetic
vortex located
at the apex of the cone.
The $d=3$ case of the Lagrangian (\ref{adss})
with $\sigma=1/2$, ${\cal M}\neq 0$, and
${\cal T}_3=0$
describes the dynamics of the charged particle near the
horizon of an extreme Reissner-Nordstr\"om
black hole \cite{kallosh,Bell}.

\section{Discussion and Outlook}

In conclusion, let us summarize
the obtained results and discuss shortly
some problems which deserve further attention.

The massless particle in
Minkowski space $\R^{1,d-1}$
can be treated as the massless particle
on the $(d+1)$-dimensional conical
hypersurface immersed in
${\cal R}^{2,d}=\R^{2,d}-\{0\}$.
In such a picture,
the rigid conformal symmetry $so(2,d)$
of the massless system in $\R^{1,d-1}$
originates from the isometry of the ${\cal R}^{2,d}$.
The reparametrization, and the local scale and
special conformal transformations
generated by the three
first class $so(2,d)$-scalar
constraints (\ref{xp2}), (\ref{xp1})
of the extended system
constitute its gauge symmetry being the
local conformal $so(2,1)$.
With this interpretation,
the nonlinear nature of the special
conformal transformations of the massless particle
in (compactified) Minkowski space
is rooted in the nonlinearity
of the map (\ref{cantr0}),
(\ref{cantr}) establishing the relation between the
phase space variables of the $d$-dimensional system
and observables of the
system on the cone.
The extended system, in turn,  can be interpreted
as the massless particle
confined to the border of the $(d+1)$-dimensional
anti de Sitter space
of infinite radius.

The rigid conformal $so(2,1)$ symmetry
is  the dynamical (hidden) symmetry of various
non-relativistic systems including the
free particle in the (punctured) Euclidean
and conical spaces \cite{Des},
the conformal mechanics model (\ref{ax}) \cite{AFF},
the charge-vortex \cite{Jackv} and the charge-monopole
\cite{Jackm,mp}
systems, and the conformal model
corresponding to the dynamics of the
charged particle near the horizon of an extreme
Reissner-Nordstr\"om black hole \cite{kallosh}.
The $so(2,1)$ and rotation symmetries
of these systems
make their dynamics
encoded in the universal Hamiltonian
structure (\ref{hamgen})
to be very similar.

By representing the Lagrangian of the
free non-relativistic particle
in $\R^1$ in the reparametrization invariant form,
we showed that the system
can be canonically related
to the massless particle model in $\R^{1,1}$.
With this correspondence, in particular, the square
of the generator of the Galilei boosts of the
non-relativistic particle is mapped into the
generator of the conformal boosts
of the massless system.
The massless particle in ${\cal R}^{1,1}=\R^{1,1}-\{0\}$
is related to the massless particle
in AdS${}_2$ by a simple canonical transformation.
As a result, the free non-relativistic particle
in ${\cal R}^1=\R^1-\{0\}$ is canonically related
to the massless particle on AdS${}_2$.
On the other hand, we showed that the
canonical transformation establishes also
the relation between the massive particle
on AdS${}_2$ and non-relativistic model
(\ref{ax}) in one dimension.
Generalizing these results,
we have demonstrated that
all the listed conformally invariant
non-relativistic $d$-dimensional systems
can be canonically related
to the relativistic particle systems
on the AdS${}_2\times$S${}^{d-1}$
described by the Lagrangian of the
universal form (\ref{adss}).

The quantization of the non-relativistic
systems presented in the reparametrization invariant
form results in the corresponding
Schr\"odinger equation.
In such a way the hidden conformal
symmetry of the equation
naturally originates
from the $so(2,1)$ symmetry
of the classical systems.
On the other hand,
conformal symmetry is
the symmetry of the
field systems appearing as a result
of quantization of the corresponding
relativistic systems described classically
by the Lagrangian (\ref{adss}).
Therefore, it would be interesting
to investigate the relationship
between the non-relativistic and relativistic
systems at the level of the corresponding
non-relativistic and relativistic field theories.
This could be helpful, in particular,
for the better understanding of the
hidden $so(2,1)$ symmetry of
some non-relativistic Chern-Simons field
\cite{conF1,conF2,dhp} and fluid \cite{hh,OR} models.
It would also be natural
to generalize the present analysis
for the conformally invariant models
of the particles with spin and
for the superconformal systems.
In particular, it would be interesting
to apply the approach developed here for the
analysis of the nonlinear supersymmetry
of the fermion-monopole system \cite{fm}.

\vskip 0.5cm
{\bf Acknowledgements}
\vskip 5mm
M.P. is grateful to D. Sorokin for useful discussions.
The work was supported by
FONDECYT-Chile (grant 1010073) and DICYT-USACH (M.P.),
and by CONICYT-Chile (C.L.).

\vskip 0.3cm
{\bf Note added in proof}
\vskip 0.5cm
After  completion and submission of this work, we became
aware of
the  two-time physics formulation of one-time systems
\cite{bda,bar},
which originates from the Dirac's analysis of the
$so(2,4)$ conformal symmetry \cite{dirac}
and to which our approach is close.
We thank  Dr. C. Deliduman  for the correspondence.

\end{document}